NUCLEAR PHYSICS & PARTICLE PHYSICS

# SUB-QUANTUM MEDIUM AND FUNDAMENTAL PARTICLES

IOAN-IOVITZ POPESCU[1], RUDOLF EMIL NISTOR[2]

[1] *Member of the Romanian Academy, e-mail: iovitz@pcnet.ro*
[2] *"Politehnica" University of Bucharest, Physics Department, e-mail: nistor@physics.pub.ro*

*Abstract*. Obtaining the rest mass of leptons with electric charge –1 is pursued by considering the existence of a medium made up of sub-quantum particles, called etherons [1, 2], having a rest energy of the order of $10^{-33}$ eV, *i.e.* at the lowest limit which is possible in the Universe. This medium is assumed to have a periodic structure, with a period of the order of $10^{-15}$ m, that generates zones of allowed and forbidden energy. The basic assumption consists in considering the photon interaction with this hypothetical medium to be similar with the interaction of the electrons with the particles of a crystalline lattice. It is further assumed that an inverse particle-antiparticle annihilation process in the presence of the periodical sub-quantum field generates the particles of the Universe. The quantization of the photons in this sub-quantum lattice is achieved with the help of the operator of the square of the energy [7] and a well-known formula of F. Bloch [5, 6] has been further used to empirically fix the lattice parameters (the etheron barrier period, width and height). The rest energy of fundamental particles would correspond to zones of allowed energy. Excellent agreement has been found with the well-known data of electronic (0.511 MeV), miuonic (105.457 MeV), and tauonic (1784.037 MeV) rest energies. The rest energies of the next heavier leptons, not yet known, are also predicted to be at 4967.500 MeV, 7840.580 MeV, and 10483.240 MeV.

## 1. INTRODUCTION

Let us start with the general question of the narrow identity of elementary particles of the same type. Why all electrons are identical, for example. For this purpose, let us consider Table 1 showing the masses of fundamental particles with spin ½ [3]. More exact values are given in [4]. It can be readily seen that in the case of quarks, while their energy increases (see the second and third generation) the interval of the rest energy within which they were observed increases too. In the following we will try to explain this remarkable property by analogy with the energy bands associated to the periodic structures of crystals. Thus, as it is well known, the allowed energy strips of particles in a periodic potential become wider

*Table 1*

Mass [MeV/c$^2$] of the three generations of fundamental spin ½ particles [1]

| Type | Q | 1 Generation | 2 Generation | 3 Generation |
|---|---|---|---|---|
| Quark | 2/3 | $u(2-8)$ | $c(1300-1700)$ | $t148000$ |
| Quark | –1/3 | $d(5-15)$ | $s(100-300)$ | $b(4700-5300)$ |
| Lepton | 0 | $\nu_e < 7,3\exp(-6)$ | $\nu_\mu < 0,27$ | $\nu_\tau < 35$ |
| Lepton | –1 | $e^-0,511$ | $\mu^-106$ | $\tau\_1784$ |

and denser as their energy increases. The main ideas of this paper are (i) the assumption that the interaction of photons with the sub-quantum medium is similar to the interaction of electrons with the particles of the crystal lattice and (ii) the fact that the known quantum particles are the result of the interaction of two photons in the presence of a sub-quantum medium according to the particle generation reaction

$$h_b \nu + h_b \nu = 2 m_0 c^2 \qquad (1)$$

(and observing the simultaneous energy, momentum, and charge conservation), where $h_b$ is Planck's constant "h bar", $h_b \nu$ is the photon energy, $c$ is the light velocity in vacuum and $m_0$ is the particle rest mass. In the following we shall presume that any elementary particle can be generated in the sub-quantum environment through the process of pair generation according to Eq. 1.

## 2. THE SUB-QUANTUM MEDIUM

Let us consider a 4-dimensional space–time cavity ($L$, $L$, $L$, $T$) containing a free particle, which is described by a Klein-Gordon steady state wave function of the form [1]:

$$\psi(x, y, z, t) = \sin(n_1 \pi x / L) \sin(n_2 \pi x / L) \sin(n_3 \pi x / L) \sin(n_0 \pi x / T) \qquad (2)$$

where $n_0$, $n_1$, $n_2$, $n_3$ are integers. The momentum components and the total energy of the particle are always subjected to the quantum conditions:

$$\begin{aligned} p_x L &= n_1 h_b / 2 \\ p_y L &= n_2 h_b / 2 \\ p_z L &= n_3 h_b / 2 \\ ET &= n_0 h_b / 2 \end{aligned} \qquad (3)$$

From this we get the quantification of momentum and energy

$$p^2 = (n_1^2 + n_2^2 + n_3^2)(h_b / 2L)^2 \qquad (4)$$

$$E = n_0 (h_b / 2T) \tag{5}$$

and the expression of the rest energy:

$$E_0^2 = E^2 - c^2 p^2 = \left( n_0^2 - \frac{c^2}{(L/T)^2} (n_1^2 + n_2^2 + n_3^2) \right) (h_b / 2T)^2 \tag{6}$$

Now, in order to ensure *the largest conceivable freedom* of the particle, the cavity will be extended to the observable Universe, obeying the cosmological relation

$$L = cT \tag{7}$$

between the Universe size ($L \cong 1.2 \cdot 10^{26}$ m) and its age ($T \cong 4 \cdot 10^{17}$ s), derived from the Hubble constant $H = 1/T = 2.5 \cdot 10^{-18}$ s$^{-1}$, where $c = 2.997 \cdot 10^8$ m/s is the light velocity in vacuum. We get in this way the quantification of the rest energy $E_0$ of any free particle in the form:

$$(E_0 / \varepsilon)^2 = n_0^2 - (n_1^2 + n_2^2 + n_3^2) \tag{8}$$

where

$$\varepsilon = h_b / 2T \cong 10^{-33} \text{ eV} \tag{9}$$

is the *etheron* energy. According to the uncertainty relation

$$\varepsilon T = h_b / 2 \tag{10}$$

the quantity $\varepsilon$ represents the smallest energy which can be measured in the age of the Universe. Thus, this ultimate minute energy appears as the lowest limit of energy scale.

The integers $n_i$ have an upper limit imposed by the following two reasons. Thus, a first condition restricts the temporal quantum number $n_0$ to values below a limit given by

$$|n_0| = 2ET/2 = E/\varepsilon \leq Mc^2 / \varepsilon \cong 10^{122} \tag{11}$$

where $M \cong Lc^2 / G \cong 10^{53}$ kg is the mass of the Universe (G = Newton's gravitational constant). A second condition confines the spatial quantum number according to:

$$(n_1^2 + n_2^2 + n_3^2) = (2pL / h_b)^2 \cong (L / L_p)^2 \cong 10^{122} \tag{12}$$

where $L_p$ is the Planck length ($L_p = (h_b G / 2\pi c^3)^{1/2} \cong 10^{-35}$ m).

Obviously, Eq. 8 subjected to conditions 11 and 12 gives all the possible rest energies $E_0$ of particles that can exist in the observable Universe. Moreover, each energy $E_0$ can be achieved in a number of states (modes), which is equal to the number of integer quadruplets $(n_0, n_1, n_2, n_3)$ satisfying Eqs. 8, 11, 12. The very small value of the rest energy of etherons given by Eq. 9 suggests a continuous energy spectrum for elementary particles, which is not the case. Consequently, the elementary particles cannot be conceived as a combination of a certain number of etherons. Alternately, in this paper we will put forward the basic assumption that etherons represent a sub-quantum medium with periodic structure within which the elementary particles can be generated. As will be shown below in Chapter 4 (c), a value $a = 0.28 \cdot 10^{-15}$ m will be taken for the (average) distance between etherons, *i.e.*, of the order of magnitude of the nucleon radius, as it has been argued elsewhere [1].

## 3. PHOTONS IN SUB-QUANTUM LATICE

In this section we will adapt the results obtained for electron energy strips in a crystal lattice to the energy strips of a sub-quantum lattice made of etherons. The allowed energy strips in the considered sub-quantum medium correspond in our vision to the rest energy of quantum particles – electrons, miuons, tauons, and so on, generated by two-photon collisions as governed by Eq. 1.

### A) THE ELECTRONS IN A CRYSTAL LATTICE

Inside a crystal lattice, the electrons move in a periodic rectangular potential field as shown in Fig. 1, where *a* is the lattice period, standing for the distance between two successive barriers, and *b* is the barrier width. This periodic potential field generates the allowed energy strips of electrons in any crystal lattice. The wave equation of the problem is given by

$$\frac{d^2\psi}{dx^2} + \kappa^2(E - U(x))\psi = 0, \quad \kappa^2 = \frac{8\pi^2 m}{h^2} \tag{13}$$

and, following Bloch [5], we require solutions $\psi$ which are periodic over a distance $L = G(a + b)$, where $G$ is a large integer. As Bloch has shown, these solutions must be of the form

$$\psi(x) = u(x)e^{i\alpha x}, \quad \alpha = 2\pi k/L, \tag{14}$$

where $k$ is an integer and $u(x)$ is a function periodic in x with the period $(a + b)$.

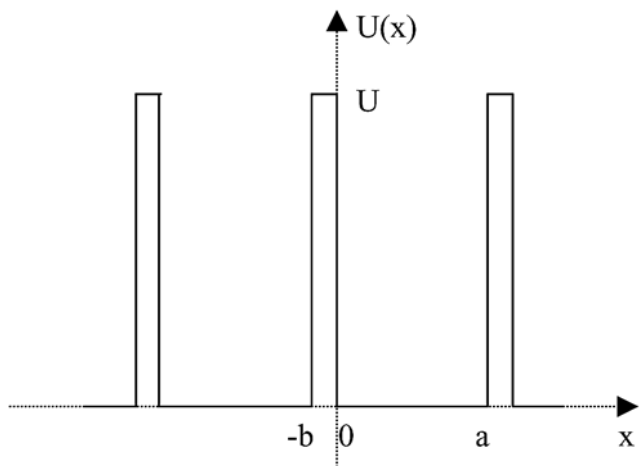

Fig. 1 – The crystal lattice potential field.

The continuity conditions of the wave function given by Eq. 14 in the leap points of the periodic potential illustrated in Fig. 1 lead to a system of four homogeneous equations which can be satisfied, as has been shown for the first time by Kronig and Penney in 1931 [6], only if:

$$\frac{\gamma^2 - \beta^2}{2\beta\gamma} sh(\gamma b)\sin(\beta a) + ch(\gamma b)\cos(\beta a) = \cos[\alpha(a+b)] \tag{15}$$

where $\beta$ and $\gamma$ are the wave numbers between two successive barriers and, respectively, within the barrier area and are respectively given by

$$\beta = \kappa\sqrt{E}, \quad \gamma = \kappa\sqrt{U - E} \tag{16}$$

Passing now to the limit where $b \to 0$ and $U \to \infty$ in such a way that $\gamma^2 b$ stays finite and calling

$$\lim_{\substack{b \to 0 \\ \gamma \to \infty}} \frac{\gamma^2 ab}{2} = P, \tag{17}$$

Eq. 15 becomes in this case

$$P\sin(\beta a)/\beta a + \cos(\beta a) = \cos(\alpha a), \tag{18}$$

a transcendental equation for $\beta a$, and finally for $E$ in terms of the space $a$ between successive potential barriers.

### B) THE PHOTONS IN A SUB-QUANTUM LATTICE

From the very beginning we have to point out the fact that the above theory of electrons in a crystal lattice is a non-relativistic one, whereas photons are relativistic particles. Therefore, in the following we will adapt the equations for electrons in crystal lattices to our hypothetical sub-quantum medium filled up with

etherons – considered to be immobile and periodically distributed in space with the period *a* to be chosen as shown in Chapter 4 (c).

We start by considering the ensemble photon – medium characterized by its relativistic particle momentum – energy quadrivector and by its total particle energy

$$E^2 = p_{0x}^2 c^2 + m_0^2 c^4, \tag{19}$$

an expression giving also a relationship between the components of the particle quadrivector. In the case of photons the second term of Eq. 19 is zero just as their rest mass is. It is important to emphasize that the photon energy remains constant in different media, though his speed and his momentum change. More specifically, the photon momentum grows and his speed diminishes for a number of times equal to the refraction index of the medium. We will further denote with index *0* the symbols in vacuum, namely

$$E^2 = p_{0x}^2 c^2 = p_x^2 v^2 \tag{20}$$

The last equation can alternately be written:

$$E^2 = p_x^2 c^2 + U_m^2 \tag{21}$$

where by $U_m$ we understand the potential of the propagation medium with respect to photons [7]. Generally, Eqs. 20 and 21 imply for $U_m$ to be an imaginary number in usual dielectrics (as far as $U_m^2 = E^2 - p_x^2 c^2 = p_{0x}^2 c^2 - p_x^2 c^2 < 0$), but a real one in waveguides or for some plasmas [8]. Using the operators $\hat{E} = E$ (the stationary case), $\hat{U}_m = U_m$ and $\hat{p}_x = -ih\nabla_x$, Eq. 21 leads us to an equation characterizing the propagation of photons along the *x*-direction, namely

$$\Delta_x \psi + \frac{1}{h_b^2 c^2}(E^2 - U_m^2)\psi = 0 \tag{22}$$

This is the stationary form of the equation of photon propagation through the considered medium (homogenous perpendicularly to the direction of propagation). For the particular case of photon propagation in the sub-quantum lattice we will consider $U_m = U$ to be a real positive number. The difference between Eq. 13 for electrons in a crystal lattice and Eq. 22 for photons in the sub-quantum lattice consists in the form of the wave numbers between successive potential barriers (β) and, respectively, within the potential barriers (γ), that is

$$\beta = \frac{1}{h_b c}\sqrt{E^2}, \quad \gamma = \frac{1}{h_b c}\sqrt{U^2 - E^2} \tag{23}$$

As will be shown in continuation below, the assumptions leading to Eq. 15 and, respectively, to Eq. 18 are strongly fulfilled also in the case of the considered etheronic sub-quantum medium characterized by the wave numbers as expressed by Eqs. 23. It is to be stressed that the quantization of the photon energy in the sub-quantum lattice has been achieved by resorting to the operator of the square of the energy, a fact meaning that the spin does not play an essential role in the considered model [7].

## 4. THE CHARACTERISTICS OF THE SUB-QUANTUM MEDIUM

Dealing with Eq. 18 requires a number of parameters characterizing the etherons (their potential height and width) and the period of the lattice they constitute, as follows:

a) ***The potential height U of the etheron.*** Due to the fact that etherons are neither the building blocks of the elementary particles, as mentioned above, nor are they affected by the presence of photons, we will assume that they possess a huge potential height, $U$, as compared with that of the photons. In this connection, the function $f(E)$ shown in Fig. 2 will be analyzed in the next Chapter 5 in its points of intersection with $f(E) = \pm 1$, Eq. 25. As it can be easily seen, the higher the $U$-potential height, the steeper becomes the slope of de function $f(E)$ and, as a consequence, the narrower will be the corresponding allowed energy band. Current measurements of the electron rest energy report a value of $m_e c^2 = 0.511212 \pm 2.2 \cdot 10^{-10}$ MeV [1], consistent with a relative error of $5 \cdot 10^{-8}$. In the following we will chose a value of the etheron potential of $U = 1.44 \cdot 10^{20}$ MeV, leading to a relative width of the allowed rest energy band of the order of $10^{-20}$. This value of the etheron potential is precisely the potential energy between two charged particles, with the same elementary electric charge and placed at a distance of one Planck length, $L_p$, one from another. It is worthwhile to say that the value of the etheron potential height $U$ chosen above is fully consistent with the "expected" zeros of the function resulting from $f(E) = \pm 1$, Eq. 25, as discussed in the final Chapter 5 below.

b) ***The width b of the etheron potential.*** As argued in [1], the etheron represents the smallest energy that can be measured in the age of the Universe and has a dimension of the order of the Planck length, $L_p$. We will consider, therefore, that the associated quantum potential of an etheron extends over a distance equal to its dimension, that is $b = L_p \cong 10^{-35}$ m.

c) ***The distance a between etherons.*** As further argued in [1], considering the minute etheron rest mass, as well as the known Universe mass and radius, it results a total number of about $10^{122}$ etherons in the whole Universe, with a density of the order of $10^{43}$ etherons/m$^3$. Consequently, the mean distance between etherons appears to be of the same order, $10^{-15}$ m, alike that of the nucleon radius [1]. In the present paper we will fit the distance between etherons in their periodic sub-quantum structure with the value $a = 0{,}28 \cdot 10^{-15}$ m.

## 5. COMMENTS AND RESULTS

We will further formally use the relationships of Chapter 3 (a) as derived for electrons in a crystal lattice and will denote the left side of Eq. 18 by $f$, namely

$$f = P \sin(\beta a)/\beta a + \cos(\beta a) \qquad (24)$$

where $P$ is given by Eq. 17. However, the crystalline lattice data will be replaced by the etheronic lattice parameters $a$, $b$, and $U$ from the above section (Chapter 4) and by the corresponding wave numbers $\beta$ and $\gamma$ for photons, Chapter 3 (b), as given by Eqs. 23. A typical behavior of the function $f(E)$ for photons in the etheron lattice is illustrated in Fig. 2, which will be further used to determine the allowed energies in the sub-quantum medium.

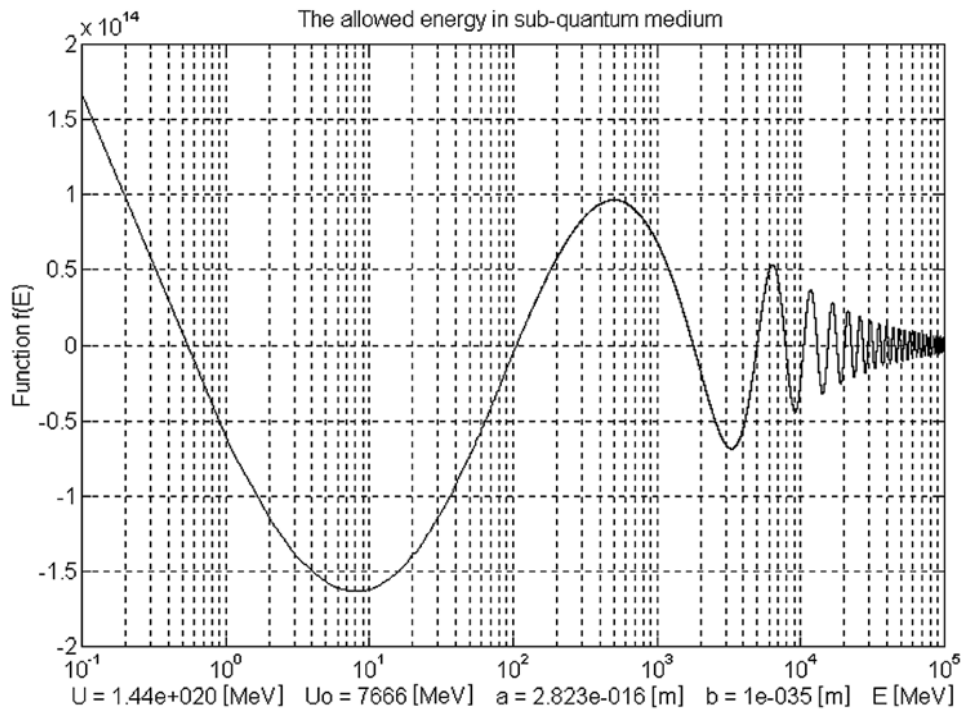

Fig. 2 – Typical behavior of the function $f(E)$.

Generally, it can be observed that when the slope of the function $f(E)$ gets steep enough in the range limited by ±1, the corresponding allowed bandwidth narrows so much, so that it could be quite well approximated by a well-defined energetic level. This is the way the present model explains why all electrons (and other leptons as well) appear to have the same rest mass.

It is further well known that the electrons can appear by a pair generation process, Eq. 1, in which the conservation of energy, momentum, and charge have to occur simultaneously. Such a reaction never could take place if the energy of the two-photons is smaller than the rest energy of the generated electrons, but could take place when it is higher, the excess energy being converted into kinetic energy of the reaction products. This can be explained if it is accepted that the first allowed energy band associated with the hypothetical sub-quantum etheron environment corresponds to the rest energy of the electrons. Generally, the function $f(E)$ vary with the energy as shown in Fig. 2 and the strips of allowed energy are those for which this function $f(E)$ takes values between +1 and –1. When the periodic potential "teeth", Fig. 1, become very high and very narrow (such as a periodic Dirac delta-function) the function described by Eq. 24 becomes periodic with a period independent of energy. In this limiting case, the solutions of the equation

$$f = P\sin(\beta a)/\beta a + \cos(\beta a) = \pm 1 \tag{25}$$

$$(n-1)\pi + 2\alpha(a+b) < \beta a < n\pi \tag{26}$$

do not satisfy the requirements needed by the rest energy of leptons inasmuch as their rest energies are not multiples of each other. However, we succeeded to obtain the masses of leptons with charge –1 as solutions of Eq. 25 by empirically introducing a dispersion law for the lattice parameter $a$ with the help of the following replacement

$$a \to a\sqrt{1+(U_0/E)^{1.74}} \tag{27}$$

with $U_o = 7666$ MeV. This correction simply means that, actually, the distance $a$ between the walls of the potential well depends on the particle energy and, specifically, the periodicity of the function $f(E)$ in Eq. 24 increases with energy. This effect can also be interpreted as a guiding effect increasing with energy, superimposed over the periodicity effect of the infinite etheronic structure. Obviously, other equivalent parametric approaches are possible as well instead of the above substitution, Eq. 27, as, for instance, to let the distance $a$, the potential $U_o$, and the power 1.74, as free parameters to be determined from the actual rest masses.

The behavior of the function $f(E)$ in various relevant intervals is further shown in Fig. 3 (between 0–2000 MeV), Fig. 4 (between 0.2–1 MeV), Fig. 5 (between 100–110 MeV), and Fig. 6 (between 1780–1790 MeV), and the comparison between experiment and theory for the rest mass of leptons with –1 charge is given in Table 2, where $\Delta E_{theor}$ is the computed allowed energy bandwidth. As seen, the agreement with extant data in literature [4] is within a relative error of only a few tenths of percents. Moreover, rest masses of trans-tauonic heavy leptons, denoted here by $\lambda_4$, $\lambda_5$, $\lambda_6$, a.s.o. can be predicted. The encouraging results, obtained with this simple model of an etheronic sub-quantum medium, hold the promise of a new perspective in the physics of sub-quantum medium and fundamental particles.

*Table 2*

Comparison between experiment and theory for the rest mass of leptons with –1 charge

| Generation | Particle | $E_{theor}$ [MeV] | $\Delta E_{theor}$ [MeV] | Rest mass after [2] [MeV] | Relative error |
|---|---|---|---|---|---|
| First | Electron e | 0.511 | 2.28e-14 | 0.5112 | 0.25% |
| Second | Muon μ | 105.457 | 2.00e-12 | 105.70 | 0.21% |
| Third | Tauon τ | 1784.037 | 2.40e-11 | 1776.99 | 0.42% |
| Fourth | heavy lepton $\lambda_4$ | 4967.500 | 2.50e-11 | | |
| Fifth | heavy lepton $\lambda_5$ | 7840.580 | 3.50e-11 | | |
| Sixth | heavy lepton $\lambda_6$ | 10483.240 | 4.00e-11 | | |

*Acknowledgments.* The authors would like to thank Dr. Nicolae Ionescu-Pallas, former senior scientific researcher at Institute of Atomic Physics, Bucharest, for his valuable comments and encouragement to publish the present paper.

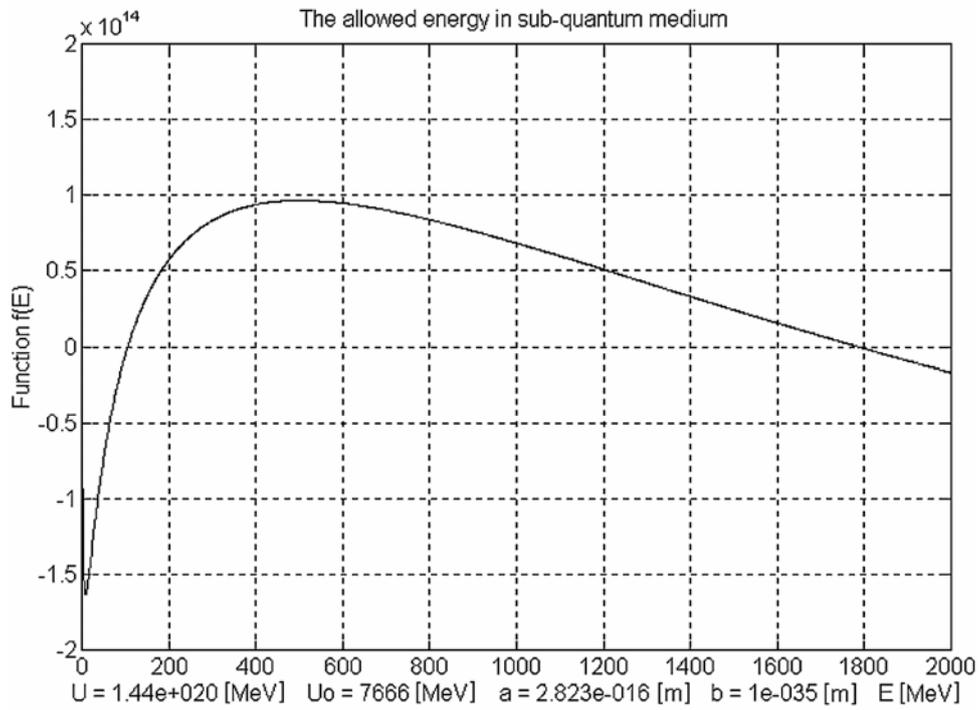

Fig. 3 – The behavior of the function $f(E)$ between 0–2000 MeV.

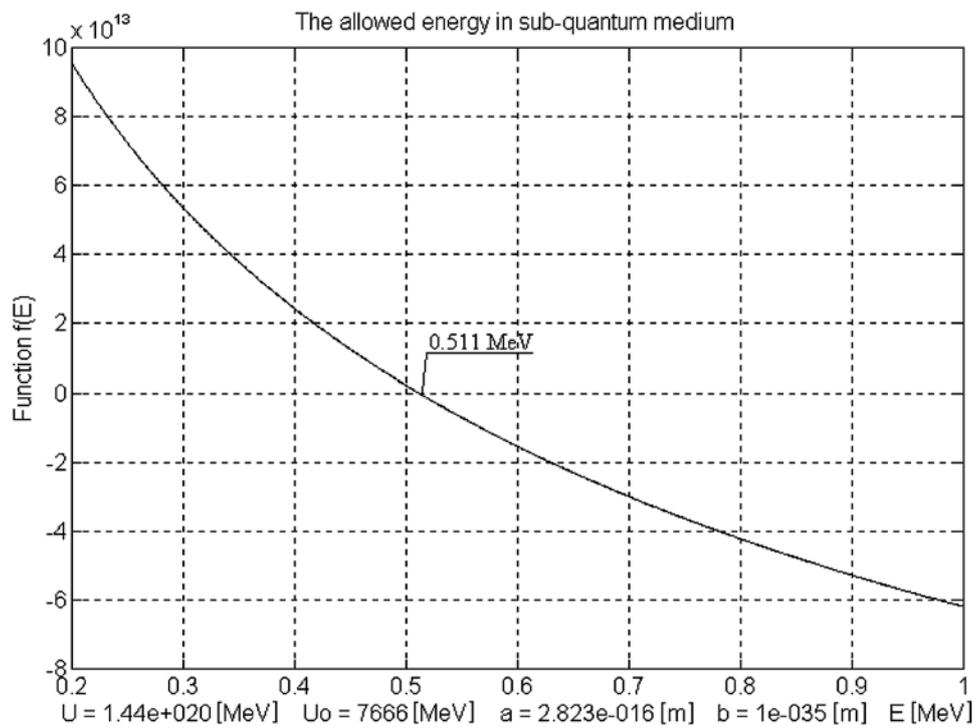

Fig. 4 – The behavior of the function $f(E)$ between 0.2–1 MeV (fixing the electron rest energy at 0.511 MeV).

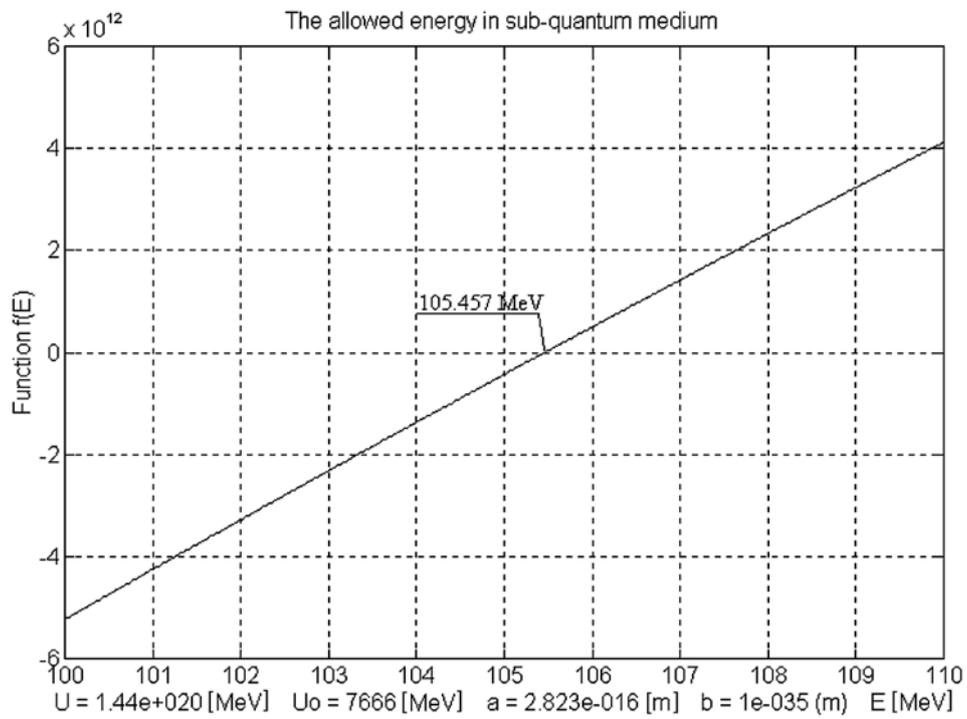

Fig. 5 – The behavior of the function $f(E)$ between 100–110 MeV (fixing the muon rest energy at 105.457 MeV).

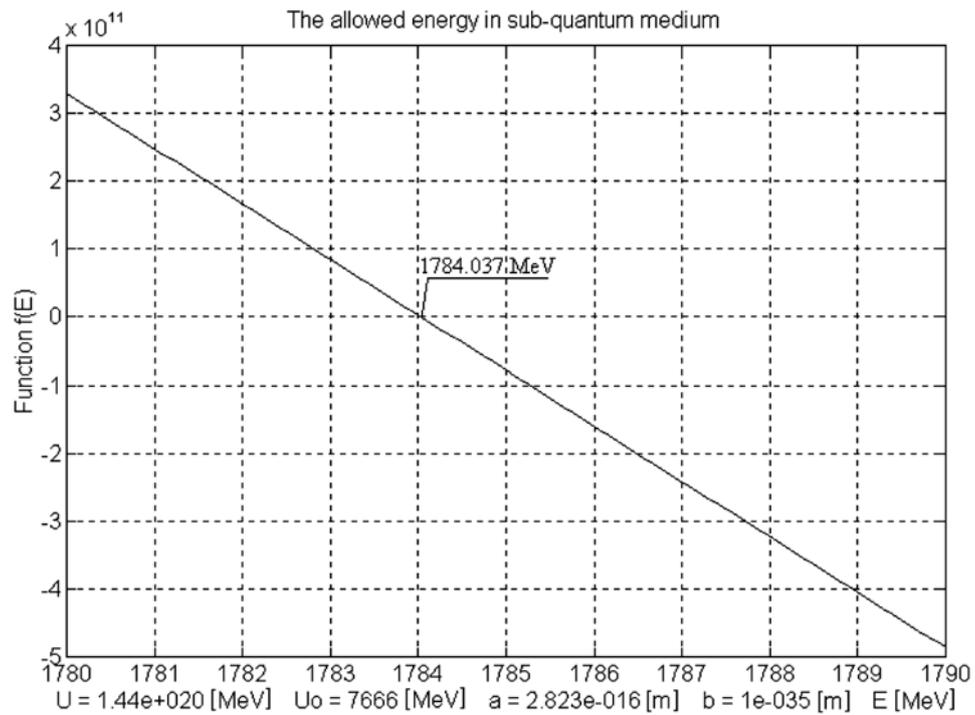

Fig. 6 – The behavior of the function $f(E)$ between 1780–1790 MeV (fixing the tauon rest energy at 1784.037 MeV).